# Mastering Rheology: A Strategic and practical guide for empowering all users

Khushboo Suman*

Department of Chemical Engineering, Indian Institute of Technology Madras, Chennai 600036 India

*Corresponding author: ksuman@iitm.ac.in

## Abstract

Rheology, the study of flow, plays a vital role in diverse industries such as pharmaceuticals, cosmetics and food. In this work, we provide a comprehensive introduction to fundamental rheological experiments and offer a strategic approach to understand the rheological behavior of any soft condensed material. We emphasize the importance of design of a good rheological experiment, input parameter and analysis of the obtained output parameter. Through the standard design of the experiment and systematic conduction of rheological experiments, we can estimate a broad range of rheological parameters such as viscosity, modulus, stability, yield stress, nonlinear behavior of the material. Furthermore, we also discuss methods for detecting and good practices for reducing some of the common experimental errors that may arise. We also present a comprehensive range of advanced rheological characterization techniques that can be pursued to gain deeper insights into the mechanical behavior and structural evolution of soft materials. Through this work, we attempt to make rheology more accessible to researchers from various domains to incorporate rheology into their characterization study and enable them to confidently navigate rheological studies and contribute to material development.

# Introduction

Rheology is the study of the deformation and flow behavior of materials.[1] It helps us understand the response of the material upon application of stress or strain, whether they behave like elastic solids, viscous liquids, or something in between (viscoelastic).[1] Usually we refer to materials as complex material and rheology of complex materials is commonly perceived as a complicated and mathematically challenging field. However, rheology is an extremely essential technique in understanding the behavior of non-Newtonian fluids and complex materials.[2–5] An understanding of the rheological behavior of the material serve as a crucial information in designing and optimizing the properties of the products used in varieties of industries including food, pharmaceutical and manufacturing.[6–11] Furthermore, rheology is often utilized as a technique to solve practical issues related with material properties. For instance, in case of any stability or textural issue with products such as ketchup, paints, personal care products, viscosity and modulus measurements help in solving the issue.[12,13] However, it is important to note that these rheological properties are not a constant and depends upon multiple parameters including shear rate, frequency, temperature.[14] Therefore, in order to gain a deeper understanding of the material, it is essential to follow a systematic approach to conducting rheological measurements, ensuring accurate and reproducible results.

Although there are many researchers working in the domain of materials who are not experts in rheology, they still employ rheological principles for enhancement of product design. For instance, to quantify the correct consumer feel of products such as peanut butter and hand sanitiser, viscosity measurements are performed at different shear rates mimicking the actual response of the material under operational shear rate range.[15,16] In the context of sustainability, rheology is increasingly perceived as important characterization technique for optimizing manufacturing processes, reducing waste, and improving the performance of eco-friendly materials. Rheology also plays an important role in additive manufacturing by providing essential insights into the flow behavior of the printable materials. [17,18] It is often required to add a binder or adjust the time of solidification of the material and rheological characterization can be very beneficial in achieving the desired viscosity for smooth flow. Furthermore, rheological characterization is instrumental in optimizing the processing parameters such as curing time and temperature to enable the design of advanced materials with desired mechanical and functional properties. Based on the wide applicability of rheology in multiple sectors spanning science fields, industry and everyday life, it becomes important to have a list of preliminary experiments to be performed to gain insight about the material behavior.

One of the major challenges is in setup of an appropriate test with well-defined parameters. Many researchers who are interested in performing rheological experiments often rely in the literature and try to replicate their experimental protocol, especially when they are new to the field of rheology. As a result, the researchers often tend to replicate the same set of experimental protocol. However, while replication of experimental protocol offers a starting point, it has its own limitations. Every material behaves uniquely and therefore setting up an experiment requires specialized knowledge of sample, range of parameters such as shear rate, stress, temperature and measurement geometry. Any change in the experimental parameters can lead to difference in results.[19] Furthermore, the literature mostly provides information about what experiments worked and gave good results. However, it often ignores details about potential issues during the experiment setup and the aspects that researchers must be careful. While there are several papers highlighting the applicability of rheology, there is a dearth of literature on systematic approach regarding conducting appropriate rheological experiment.[12,13]

A few tutorials have recently been published on performing rheological experiments. However, they are focussed to a certain class of materials such as polymers[20,21] and pharmaceuticals[22]. Simoes *et al.*[22] presented a methodological tutorial for rheological analysis using a 1% hydrocortisone cream as a model formulation. By assessing the impact of critical method variables on key rheological parameters, the work presented a framework for measuring oscillatory response and thixotropy. Ricarte and Shanbhag[21] presented a review on linear rheology of covalent adaptable network systems. They explain fundamental concepts such as relaxation spectra and time–temperature superposition, and offers best practices and models for analyzing the rheology of covalent adaptable network systems. Furthermore, Ewoldt *et al.*[23] highlighted the common difficulties in assessing shear material properties for soft, and living biological fluids, and discusses techniques to minimize experimental errors, using two systems including an aqueous polymer/fiber network and a suspension of microalgae.[23] While the review of Ricarte and Shanbhag[21] covers some standard linear rheological tests targeted for covalent adaptable networks, the article by Simoes *et al.*[22] presents the risk associated with many rheological variable on the data analysis. Furthermore, Ewoldt *et al.*[23] address how real-world limitations distort rheological data and present case studies using ultra-soft biomaterials. While all of these works contributes meaningfully to the rheological literature, there is an opportunity to present a broader and more integrated view of rheology and its utilization across different varieties of materials. Also, given the limited number of papers focused on standardizing rheological experiments, there is a critical need for a comprehensive review that outlines essential experiments and a practical guideline for obtaining meaningful insights into the flow behavior of soft materials.

This study provides a roadmap of rheological experiments to perform when we begin with any new and/or unknown material sample. Since material attributes govern the rheological products, a guide to rheological experiment will be beneficial in shortening the time period of product development and improve the product quality. Key material attributes which can be accessed using rheology includes viscosity, modulus, yield stress, long-term stability. Importantly, all the mentioned parameters are not a constant and is defined at particular process parameters including shear rate, temperature or frequency. Rheology can be applied at various stages of new product development, from the initial composition design to the final product quality control. When preparing a dispersion, measuring its viscosity ensures it meets the desired viscosity for its intended use and provides a satisfactory consumer experience. The knowledge of viscosity also helps in determining if a rheology modifier is needed to achieve the desired consistency. Furthermore, in many industrial applications, it is desirable for materials to exhibit shear-thinning behavior, where the flow becomes easier or smoother under shear force. This property is particularly useful for applications such as material pumping and products applied to the skin.[24–26] In addition, it is also beneficial to have a certain yield stress so that the material flows only when the applied stress is above the yield stress value.[27] Once the desired viscosity and consistency of the formulation are achieved, the final and crucial step is to perform a stability analysis. Rheology can help in understanding the stability of the material by measuring time-evolution of its properties as a function of varying temperatures so as to understand the performance across different weather conditions.[28] Once these parameters are understood under various processing conditions, it provides a deeper insight into the product, and this rheology-based approach can significantly accelerate product development.

In this work, we outline a comprehensive framework for conducting rheological characterization of any material. We provide best practices to be followed in designing a rheological experiment and how to interpret the obtained results. We also discuss several common artifacts that can lead to inaccurate results, explaining how to identify them and implement appropriate measures to eliminate them. We also suggest changes that can be implemented in the experimental protocol based on the prior knowledge of the material. We hope to provide a comprehensive roadmap to perform rheological characterization of any material system and help them better design the experiments. We hope this article will welcome many soft material researchers to utilize rheology in understanding the viscoelastic property.

## Materials and Methods

In this review, we present results using schematics and also by performing actual rheological experiments on two synthetic material systems which are widely used in industries as rheological modifier. These model systems have been extensively investigated due to their well-characterized structures and versatile applications, making them ideal candidates for illustrating some of the fundamental rheological concepts. The first material is a nanoclay dispersion of Laponite® XLG (a registered trademark of BYK Additives). It is a synthetic clay particle of around 30 nm in diameter and 1 nm in thickness. The nanoclay dispersion is known to undergo a sol to gel transition spontaneously with time. The second material system explored in this work is Pluronic F127, a triblock copolymer purchased from Sigma-Aldrich. The Pluronic solution is known to form a thermoresponsive system which form a soft solid state upon increase in temperature. An elaborate discussion about the material and sample preparation protocol can be found elsewhere. [29,30] In this study, 2 weight % nanoclay dispersion (Laponite® XLG with 4 mM NaCl) and 20 weight % triblock copolymer (F127) solution has been used. All the experiments were performed on Anton Paar 502 rheometer using cup and bob geometry. The details of the different rheological experiments performed are described in each experiment type and listed in table.

## Rheological Measurements

Most materials that we commonly encounter in our daily life, industry or nature are neither perfectly elastic nor perfectly viscous in nature. They are intermediate between elastic solids and viscous liquids and are called as viscoelastic materials. The fundamental theoretical aspects of rheology and the governing equations are universal to all viscoelastic materials and are well described in standard books on Rheology.[31,32] To perform any rheological experiment, the input parameter is typically either stress, strain or shear rate. These rheological variables get interpreted from the raw machine measurements of torque, angular displacement and angular velocity. Few of the important rheological parameters which help in describing the response of the soft material are viscosity, elastic modulus, viscous modulus, damping factor ($\tan \delta = G''/G'$) and yield stress. The variation of these rheological parameters provides insight into the behavior of the material and is discussed experiment wise below. In Figure 1, we present a comprehensive roadmap to perform rheological characterization. Figure 1 briefly summarizes the input variables, output variables and the inference we can draw from each of the rheological experiment to have an enhanced understanding of the material behavior. We also present a summary of the key input and output parameters, typical representation of input function and material response behavior as well as the main inference drawn from the particular experiment in Table 1. In this section, we list the details of some of the most fundamental rheological experiments to be performed on any soft material. We discuss the design of the experiments comprehensively with all the input parameters and suggest a good range of values to be utilized for the experiments. We also the discuss the modifications and important aspects to consider while designing the experiment as per the material under study.

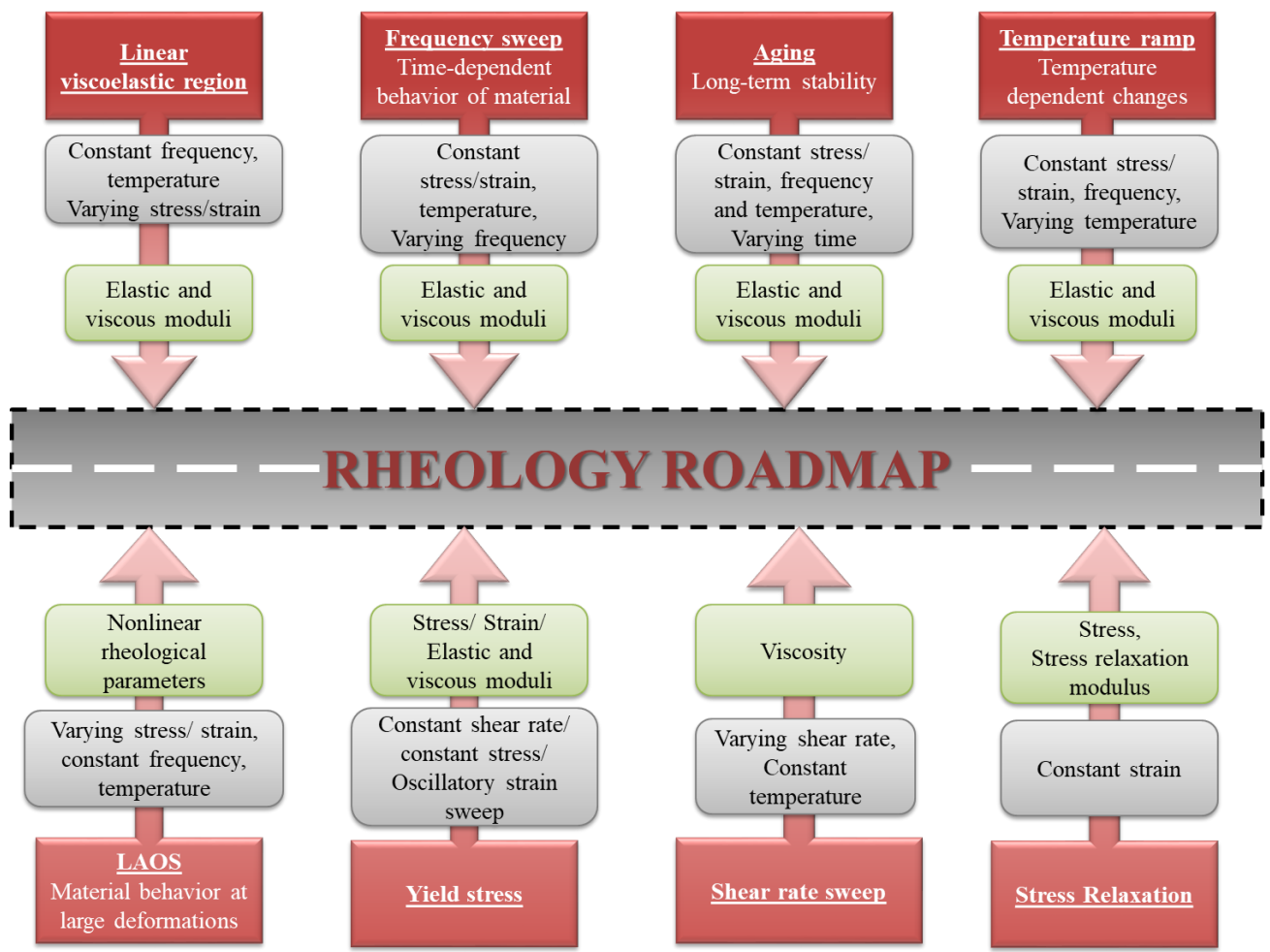

**Figure 1.** An overview of rheological experiments. The grey box mentions the input parameter and the green box mentions the output variable obtained from that particular experiment mentioned in red box.

Linear viscoelastic region

To perform any rheological experiment, first and foremost it is important to identify the linear viscoelastic (LVE) region. In the LVE region, the material properties such as modulus and viscosity are independent of the magnitude of the applied deformation. Therefore, application of deformation within the LVE region does not alter the intrinsic microstructure of the sample. All the reported material properties such as, elastic modulus, viscous modulus, complex viscosity, gel strength, tan $\delta$ are usually reported within LVE region unless mentioned otherwise. Therefore, estimation of LVE region is crucial in characterizing the intrinsic behavior of the material and designing any further rheological experiment to compute its material properties. Mathematically, LVE region in an oscillatory flow field is characterized by linear relationship between oscillatory stress and strain where the proportionality constant is complex modulus.

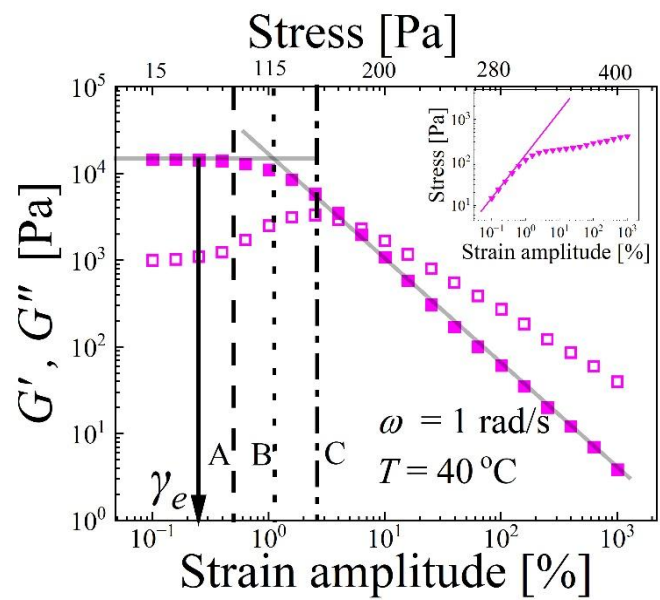

**Figure 2.** The variation in elastic ($G'$, closed symbols) and viscous ($G''$, open symbols) modulus is plotted as function of oscillatory strain for a triblock copolymer solution at 40ºC. The corresponding oscillatory stress in the material is reflected in the top x-axis. The lines denote the construction utilized for determining yielding of the material which is discussed in the yield stress section. The inset shows the variation in oscillatory shear stress as function of oscillatory shear strain for the same data as in the main figure. The solid lines denote a linear fit to the data.

The most popular way of identifying the LVE region is by conducting an oscillatory strain sweep experiment wherein the strain amplitude is increased in a logarithmic fashion from a very low value (~0.01%) to high value (~1000%) at a constant angular frequency and constant temperature. The lower limit of amplitude of applied strain can depend on the specification of the instrument being employed. At very low strain amplitudes, noisy data may appear which suggests that the measurement is approaching instrument limits. The angular frequency value is so chosen that the data point can be collected in reasonable experimental timescale without the sample changing its microstructure inherently during the sweep (can be 1 or 10 rad/s). It is a good practice to conduct the strain sweep in increasing manner because application of very high strain will disrupt the microstructure and it may take some finite time for the material to regain its microstructure, if recovery is possible at all. The elastic and viscous moduli is plotted as a function of oscillatory strain in Figure 2. The results are most effectively presented in a double logarithmic plot of the elastic modulus ($G'$) and viscous modulus ($G''$) as a function of oscillation stress. It can be seen that the moduli value remains constant until a specific value of strain beyond which the moduli starts to change with strain applied. The region where the moduli remains constant is classified as the linear viscoelastic region. The linear viscoelastic region has no lower limit but it is bounded by an upper limit. In order to identify the critical strain beyond which the moduli are no longer a constant, there have been a couple of approaches in the literature. Some of the most common techniques of identifying the critical strain (limit of LVE region) includes the point at which there is a 5% drop in modulus value (marked by line A in Figure 2), or intersection of low and high strain asymptotes (marked by line B in Figure 2) or the point of maxima in viscous modulus (marked by line C in Figure 2) [9,33]. After identification of the LVE region, it is important to choose a particular value of strain at which further rheological experiments can be performed. In principle, the material property is independent of magnitude of deformation field within the LVE region. However practically, measuring the properties in a well-defined range of the instrument specifications is recommended to ensure good quality data. Furthermore, choosing a strain value too low may lead to noisy data. This aspect is discussed in detail in Temperature ramp section when the same experiment is conducted using different value of deformation field in LVE region. Therefore, a strain value close to the critical strain value is denoted as $\gamma_e$ a good starting point to use as standard value of deformation field to be employed in other rheological experiments. The corresponding value of stress at $\gamma_e$ can be taken as the operational stress value to conduct further experiments. Therefore, in order to conduct any subsequent rheological experiment which is outlined below or beyond, one can either choose to apply a strain or stress input estimated from the LVE region analysis.

Another way of estimating the linear viscoelastic region is by plotting oscillatory stress as a function of strain as shown in the inset of Figure 2. The stress – strain curves for the data shown in Figure 2 is plotted in the inset for the triblock copolymer solution. It can be seen that stress increases linearly at strain amplitudes less than critical strain. Beyond the critical strain, the dependence of stress on strain deviates from linearity. Therefore, the plot of oscillatory stress to strain can also be used to characterize the LVE region.

In case we have a temperature dependent system, it is advisable to conduct the linear viscoelastic measurement at temperature where the system is most fragile. Therefore, any strain value in the LVE region of a fragile state will certainly remain within the LVE region when the material strengthens, solidifies or consolidates. There exists a wide range of polymer and particulate dispersion which exhibits thermoresponsive behavior wherein it can either form a soft solid with either increase in temperature or decrease in temperature.[14,34] Owing to this variability, when working with an unknown sample, it is advisable to perform strain sweep experiment at the lowest and highest temperature within the intended experimental temperature window. This approach helps in estimating the critical strain values at which the material maintains linear viscoelastic behavior, thereby ensuring that reliable and high-quality rheological data can be obtained consistently across all temperatures of interest. In Figure 3, we plot the variation of $G'$ and $G''$ as a function of strain amplitude for a thermoresponsive system of triblock copolymer. The corresponding stress-strain curves in shown in the inset of Figure 3. The polymer solution is known to undergo sol to soft solid transition as a function of temperature.[30,35] Therefore, we conduct strain sweep experiment at various temperatures in the temperature window where it undergoes the transition. It can be seen that at low temperature, the value of moduli is quite low owing to the liquid like state of the material. The liquid like behavior of the polymer solution is also highlighted by the linear dependence of stress on strain in the inset. It can be seen that the linear dependence shown by black line in the inset extends through the entire range of strain explored at 10ºC. Upon increase in the temperature, the amount of stress developed in the material and correspondingly the moduli increases. Importantly, the LVE window is not identical at all temperatures and shifts to lower values upon increase in temperature. At 30ºC and 40ºC, the moduli curves are almost on top of each other as the microstructure remains almost similar once the system has transformed to a soft solid state. Therefore, we should select the critical strain value from the moduli curves corresponding to 40ºC for conducting any rheological experiment in the temperature window of 10 ºC to 40 ºC such that the sample is always in LVE region.

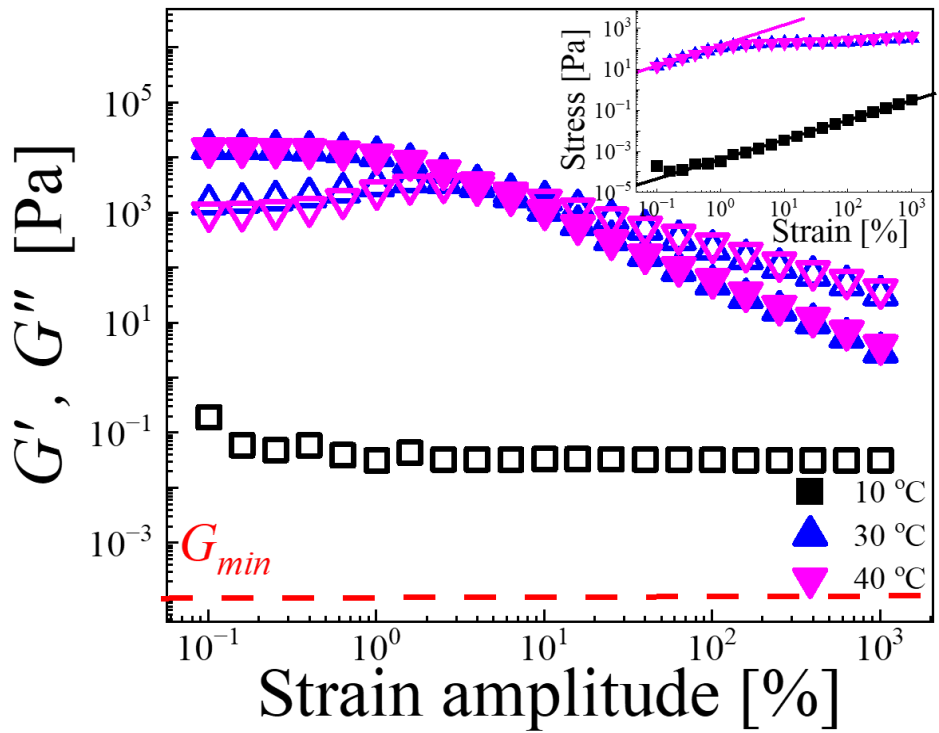

**Figure 3.** The variation in elastic ($G'$, closed symbols) and viscous ($G''$, open symbols) modulus is plotted as function of oscillatory strain for a triblock copolymer solution. The measurement is performed at 10ºC, 30 ºC and 40ºC corresponding to liquid-like, transition state and soft-solid state. The dashed line indicates the lower limit of measurement in modulus ($G_{min}$). The inset shows the variation in oscillatory shear stress as function of oscillatory shear strain for the same data as in the main figure. The solid lines denote a linear fit to the data.

Frequency Sweep

Having identified an operational strain (or stress) value belonging to the LVE region, the next important experiment to conduct is frequency sweep experiment. A frequency sweep provides insights into the material's time-dependent behavior. The data at high frequency talks about materials response at short timescale whereas low frequencies corresponds to response at long timescales or at rest. A typical frequency range from 0.01 to 100 rad/s is chosen for conducting the experiment with logarithmic increase in the frequency. However, for a stress-controlled instrument, inertia begins to play a role at higher frequency which is characterized by steep increase in the modulus value. The inertia dominated results can also be identified using the raw phase angle measured by the rheometer. Generally, raw phase greater than 170º is associated with inertia dominated region. One must be careful in identifying the frequency range where inertia dominates and the data in that region should not be accounted as part of the material response. The influence of inertia can be minimized by using a light-weight geometry such as those crafted using aluminium. Another technique to reduce the inertial effects is by increasing the sample stiffness which can be achieved by reducing the gap thickness.[23] A typical response of the frequency sweep behavior is shown in Figure 4. The dependence of elastic and viscous modulus as a function of frequency gives information about the viscoelastic behavior of any material. In Figure 4, we plot the variation of the viscoelastic moduli as a function of angular frequency at different times since sample preparation. The nanoclay dispersion is a spontaneously gelling system and therefore different time instances correspond to different microstructural arrangement of the clay particles leading to different states. At early times, $G'$ is below the measurable limit of the instrument while $G''$ is measurable, suggesting liquid-like behavior. As time progresses, the value of $G'$ gains magnitude and can be seen to crossover $G''$. The crossover suggests a transition from viscous to elastic behavior as the frequency increases. Importantly, the inverse of crossover frequency leads to the characteristic relaxation timescale of the material at that particular state. Upon increase in time, the value of $G'$ surpasses $G''$ and shows a weak dependence on frequency. Such frequency independent behavior with $G'$ greater than $G''$ is suggestive of a gel like structure in the material. Another important parameter which we get from conducting frequency sweep experiment is complex viscosity ($|\eta^*|$) which is plotted in Figure 4b. The magnitude of the complex viscosity and its dependence on frequency gives a comprehensive understanding of material's overall resistance to flow. At early times, the value of $|\eta^*|$ remains constant as a function of frequency suggestive of a liquid-like behavior. As time increases, the nanoclay dispersion transition from a sol to gel-like structure which is marked by the increase in $|\eta^*|$. Furthermore, $|\eta^*|$ can be seen to decrease with increase in the applied frequency suggestive of a shear thinning behavior of the material. The value of $|\eta^*|$ is very crucial in leading to the estimation of other rheological parameters such as zero-shear viscosity, equilibrium modulus and largest relaxation time of the material.[29]

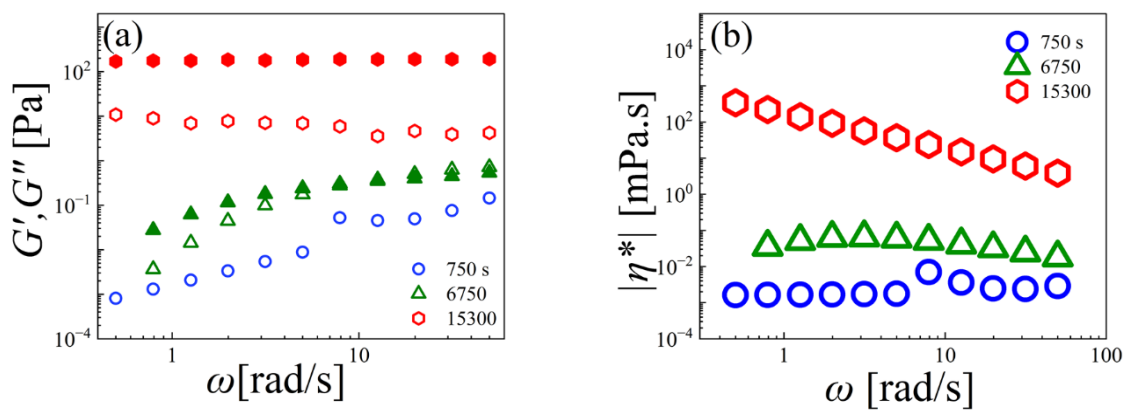

**Figure 4.** (a) The variation in elastic ($G'$, closed symbols) and viscous ($G''$, open symbols) modulus is plotted as function of angular frequency at different times since sample preparation for a colloidal dispersion of nanoclay at 28ºC. The dashed lines denote the lower limit of measurement in modulus value. (b) The corresponding value of complex viscosity ($|\eta^*|$) is plotted as function of angular frequency at identical times in Figure a.

For materials undergoing some slow changes with time such as gelation, polymerization, one can perform time-resolved rheometry.[36] In this technique, we subject the sample to multiple cycles of frequency sweep, one after another. As the structural changes progress, the dependence of the moduli on frequency will evolve with each cycle. By comparing the moduli-frequency dependence at different time intervals, one can gain a clear understanding of the extent of gelation or other structural transformations occurring in the sample. For a system undergoing liquid to gel transition, the point of critical percolation can be identified by the identical power-law dependence of elastic and viscous and moduli on frequency.[29] For a sample exhibiting changes with temperature, it is also possible to conduct cyclic frequency sweep experiment with continuous change in temperature.[30,34]

There can be another class of material which undergoes rapid changes in properties.[37,38] In that case wherein the material is evolving very fast with time, it may be challenging to obtain a comprehensive understanding of the material's response from a single frequency sweep, as the sample undergoes significant structural changes within the sweep itself. For such rapidly mutating samples, employing optimally windowed-chirp (OWCh) sequence developed by McKinley and coworkers can be extremely beneficial. [37,38] OWCh sequence involves application of frequency which changes continuously over time. This gradual variation in frequency allows chirps to effectively probe the material's response over a broad range of frequencies within a single sweep. Therefore, OWCh is a beneficial technique for gaining insights into the structural changes of a sample in a shorter duration of time.[37,38]

While the frequency sweep response offers valuable insight into a material's viscoelastic behavior, but its utility is limited by the accessible frequency range of the rheometer which covers 3-4 decades in frequency. Therefore, we do not get an idea about the complete viscoelastic spectrum of the material by conducting experiments in a limited range of frequency.[39] For thermorheologically simple materials, it is possible to get a viscoelastic spectrum expanding several decades of frequency outside the instrument range by performing experiments at varying temperature. By thermorheologically simple, it means that the sample does not undergo any phase change in the interested temperature range. We, therefore, conduct experiments at varying temperatures but identical frequency window, and then simply shift the curves to generate a master curve. This is known as the time-temperature correspondence.[39] A schematic representation of the superposition of the moduli curves to obtain a master curve is shown in Figure 5. The master curve provides a comprehensive view of the material's viscoelastic behavior over a wide range of time scales and frequencies, offering an extended understanding of its response across several decades of time.

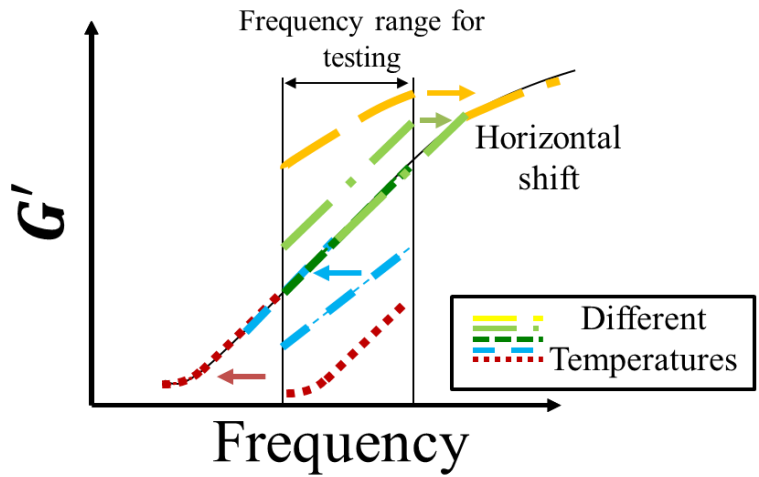

**Figure 5.** A schematic representation of time-temperature superposition. Each curve corresponds to frequency sweep conducted at different temperatures which can be shifted horizontally to obtain a master curve. The obtained master curve gives information of material behavior at varying timescales. Figure inspired from Ref [20].

For materials undergoing crosslinking, their rheological properties continually change as a material transforms from a liquid to soft solid. As a result of variation in rheological properties as a function of time, the traditional time temperature superposition cannot be applied. For such materials, the concept of effective time has been introduced where the material clock is readjusted to accommodate the time-dependent behavior. The effective time is defined by scaling real time using time dependent relaxation time. The use of effective time domain instead of real time helps in superposition of the rheological parameter and this approach is known as the effective time theory. This is an advanced form of traditional time-temperature superposition that can be applied to materials undergoing structural evolution as a function of time.[40,41]

Aging

Most of the complex materials we deal with are arrested in out-of-equilibrium state. As a result of this arrestation, the systems tries to slowly, yet continuously, evolve to a lower free energy state as they attempt to achieve equilibrium.[42] This process results in continual change in properties which is called as physical aging. An aging study gives insights about the macro- or micro-structural rearrangements occurring as a function of time and therefore, provides understanding about the stability of the sample. In the aging experiment, the sample is subjected to a small deformation field in LVE region, constant frequency and temperature and the evolution of moduli is noted as a function of time. Performing aging experiment is an accurate method of manifesting the stability of the sample rather than visual observations or viscosity measurements. The time evolution of moduli gives insights into key phenomena such as phase separation and structure formation. The evolution of viscoelastic moduli in an aging experiment on a thermoresponsive dispersion of silica particles performed by Suman and Wagner is shown in Figure 6.[14] The aging experiment was performed at two different temperatures, corresponding to two different states, achieved by thermal quench from a higher temperature. It can be seen that at 24°C, the moduli increases slowly yet continuously with time. This indicates that the material is undergoing a process that imparts stability to the dispersion. This is often seen in systems undergoing gelation, polymerization, or crosslinking, where the material develops a consolidated structure over time, improving its stability.[43–45] Upon examining the aging curve at 21°C, it can be seen that there is a sudden drop in elastic and viscous modulus. Such abrupt changes in moduli can be indicative of phase separation [14]. Therefore, an aging experiment provides deeper understanding of how its stability or structural integrity develops or deteriorates over time.

For systems which does not undergo phase transition with temperature, we can perform the aging experiment at elevated temperatures to expedite the stability analysis. Temperature plays a significant role in the aging behavior of complex materials. An increase in the temperature imparts higher thermal energy to the constituents of the sample, as a result, they exhibit accelerated aging kinetics. Such enhancement in aging behavior is commonly observed in many materials including particulate dispersions[46] and polymer solutions.[42] Particularly, aging studies is very crucial for industrial products since it directly relates with the shelf-life of the product. Therefore, one can in principle perform aging studies at elevated temperatures to get an insight about the sample's long-term stability.

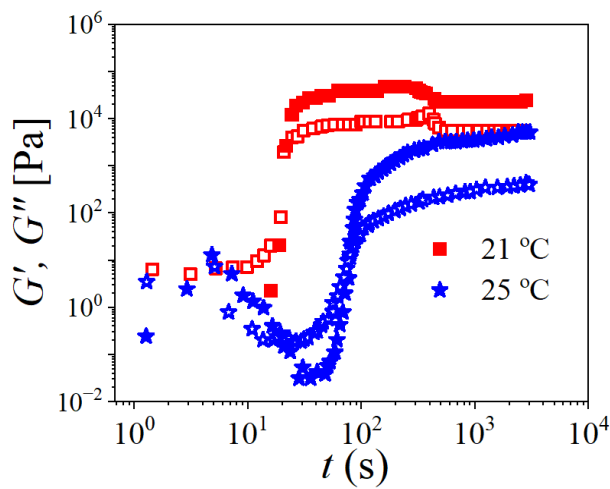

**Figure 6.** The variation in elastic ($G'$, closed symbols) and viscous ($G''$, open symbols) modulus is plotted as function of time at 21ºC (red squares) and 25ºC (blue stars). The data is taken from Ref [14].

## Temperature ramp

Temperature plays a vital role in the behavior of a complex material as we already discussed in the aging and frequency sweep sample. However, some complex materials can be extremely sensitive to changes in temperature and may undergo structural changes such as polymerization or gelation with change in temperature.[30,34] In order to examine the effect of temperature, we can perform a simple temperature ramp up and down experiment. A ramp experiment involves a continuous and smooth change in a parameter. It is important to note that contrary to ramp, a sweep protocol changes the parameter in discrete stepwise manner. In order to investigate the effect of temperature, it is usually recommended to change temperature in a ramp manner. The evolution of moduli for a polymer solution is shown in Figure 7a. The moduli changes by five orders of magnitude with change in temperature which suggest the sample undergoes a sol-soft solid transition. Furthermore, upon cooling, the system comes back to the initial sol state, however, the exact path traversed while cooling is not identical. This suggests that the material does not follow the same path at microstructural level and as result, the moduli curves do not overlap during the heating and cooling cycle. Therefore, temperature serves as a key trigger for inducing significant changes in the gel's structure, mechanical properties, and behavior. By altering the interactions between the polymer chains (such as hydrophobic or hydrophilic forces), temperature influences the gel's ability to swell, contract, change its viscosity, and undergo phase transitions between liquid-like and solid-like states. These temperature-sensitive changes are crucial for applications in drug delivery, biosensing, and other fields that require precise control over material properties in response to temperature fluctuations.

In Figure 7b, we plot the evolution of viscoelastic moduli as a temperature upon varying the magnitude of deformation applied during the experiment. For the same temperature ramp experiments, stress magnitudes of 0.1 Pa, 10 Pa and strain magnitudes of 0.1% and 0.5% have been used. As can be seen from Figure 2, all these magnitudes of deformation field lie within the linear viscoelastic region. In principle, application of deformation field within the LVE does not perturb the inherent microstructure of the material and leads to the same results. The same can be seen from Figure 7b. The value of moduli at 40°C remains the same irrespective of the deformation field applied. However, it is important to note that application of lower magnitudes of deformation field results in scattered data at low temperature as highlighted in the inset of Figure 7b. Therefore, as discussed before, it is important to select the magnitude of deformation field closer to the limit of linearity to get a good dataset.

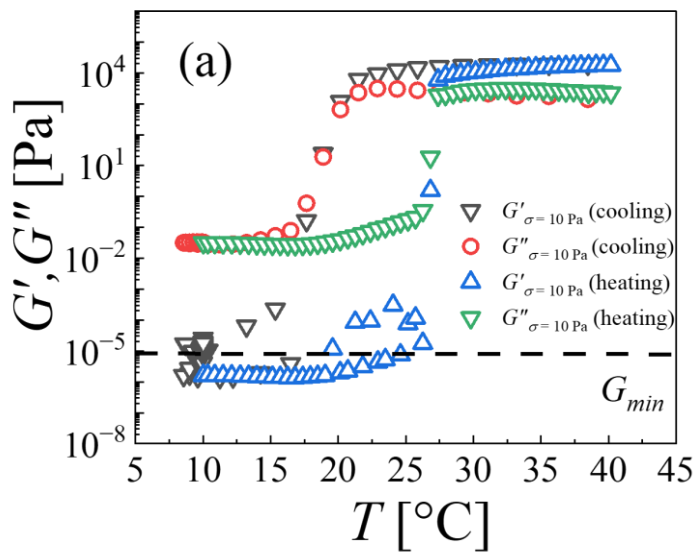

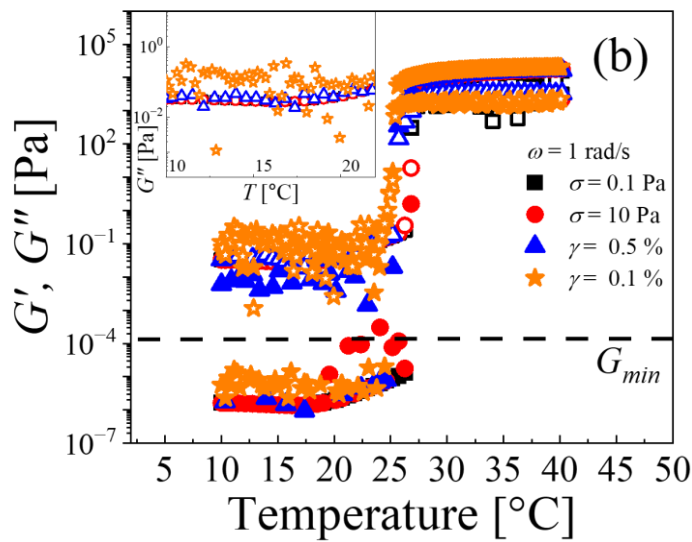

**Figure 7.** (a) The variation in elastic ($G'$, closed symbols) and viscous ($G''$, open symbols) modulus is plotted as function of temperature for a triblock copolymer solution forming thermoresponsive gel. The temperature was increased from 10ºC at a constant ramp rate of 1ºC/min. (b) The temperature ramp experiment was performed at multiple values of stress and strain, all belonging to linear viscoelastic region. The inset highlights the viscoelastic moduli data at low temperature.

Large amplitude oscillatory shear (LAOS)

The above mentioned classical rheological tests gives an understanding of the linear viscoelastic response of the material. However, under practical scenarios of industrial processing conditions, the deformation field applied to the material can be both rapid and large. Therefore, a comprehensive sample characterization requires the knowledge of material response to nonlinear flow fields as well. Large amplitude oscillatory shear (LAOS) has become an ideal method for assessing nonlinear rheological behavior, as it allows for independent control of both amplitude and frequency, enabling adjustments to both the strength and timescale of the measurements. LAOS rheological analysis of complex materials offers a more detailed understanding than traditional SAOS (Small Amplitude Oscillatory Shear) testing protocol, especially in capturing the nonlinear behavior of food materials under stress. This method helps elucidate phase transition processes and provides valuable insights into how different processing techniques impact the texture and structure of the final product.

A simple strain sweep experiment can be performed to provide insight into the nonlinear behavior of the material.[47] The typical response of a material to strain sweep in shown in Figure 8a. The high strain amplitude region where the moduli begin to exhibit a dependency with strain amplitude is referred as the nonlinear zone. In order to understand the nonlinear behavior of the material, it is important to enable record the intracycle response while doing the experiment. This involves recording the fundamental and additional higher harmonics in frequency domain. It is important to have appropriate license or active recording of higher harmonics to enable capture the raw data within a sinusoidal input of strain. The graphical representation of intracycle stress and strain is known as Lissajous-Bowditch curves (commonly also known as Lissajous curve).[47] The abundant raw data collected at each strain amplitude can be analysed using a wide range of nonlinear parameters such as Fourier transform coefficients, Chebyshev polynomials, local elastic and viscous modulus and cage modulus.[47] A comprehensive review on definition of all the mentioned non-linear parameters and their interpretation is already available in the literature.[47-49] The nonlinear data can be analyzed using open-source software tools developed by the researchers in this field worldwide such as MITlaos and SPP-LAOS (Sequence of Physical Processes-LAOS). These programs can be accessed upon reasonable request made to the respective authors.[48,50] The nonlinear parameters for the nanoclay dispersion in shown in Figure 8b, c and d. The intracycle data for specific strain amplitudes highlighted by symbols in Figure 8 a have been analyzed and the overall Lissajous plot, elastic Lissajous plot and viscous Lissajous plot in shown in Figure 8 b, c and d respectively. It can be seen that as the strain amplitude increases, the Lissajous curves deviates from a perfect ellipse. The deformation in the Lissajous curve can be further analyzed by estimating the local elastic and viscous modulus and cage modulus.[47] A quantification of nonlinear response of the material in terms of non-linear parameters material properties also renders insights about the microstructural changes occurring in the material. For instance, based on the values of non-linear moduli, it is possible to have information if the material is strain thinning or hardening while application of large deformation fields.[47]

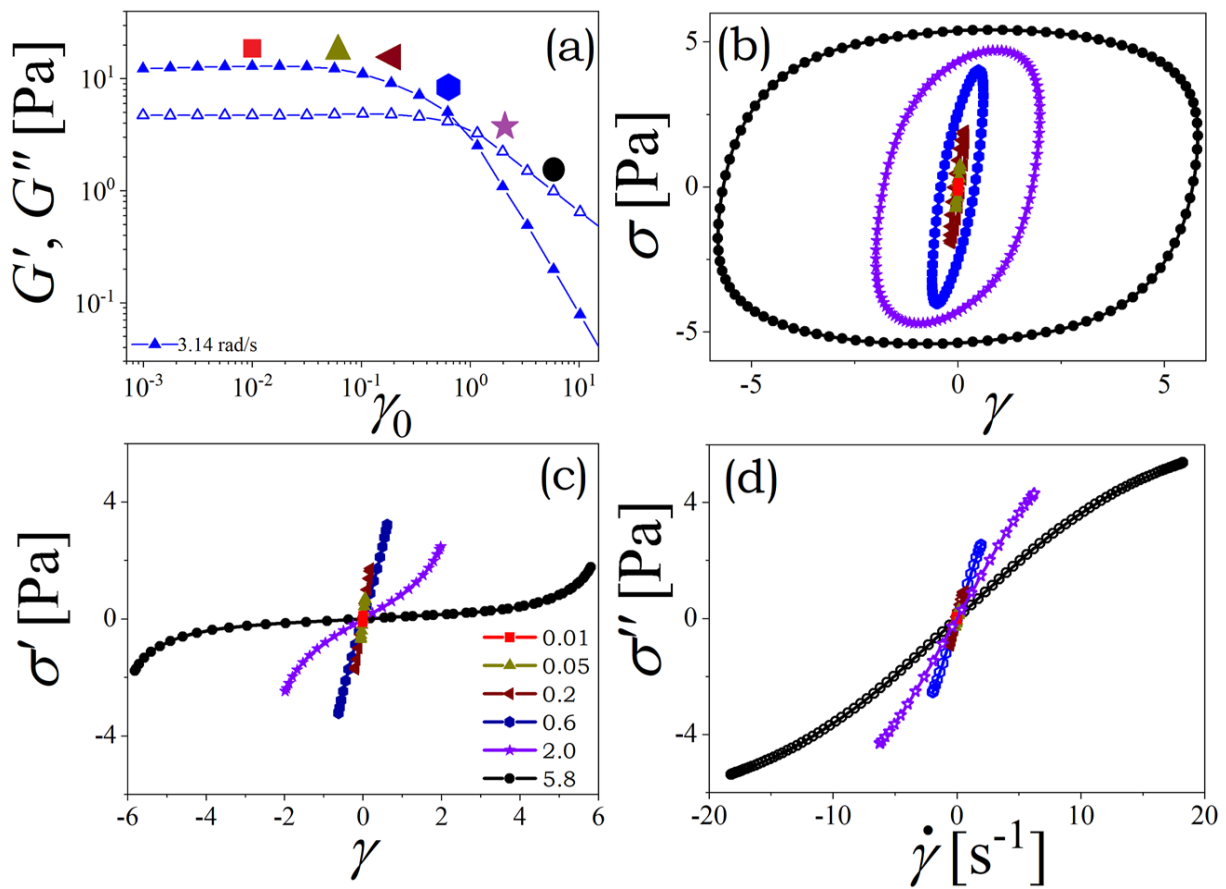

**Figure 8.** (a) The variation in elastic ($G'$, closed symbols) and viscous ($G''$, open symbols) modulus is plotted as function of strain amplitude in a strain sweep experiment. The intracycle response at fixed strain amplitudes (shown by different coloured symbols in Figure 8 a) is demonstrated in b, c and d. The total stress, elastic stress and viscous stress are shown in b, c and d respectively. Reprinted from Ref.[49] with the permission of AIP Publishing.

## Yield stress

Yield stress is an important parameter for many complex materials which provides useful information about the stability and shelf life of the product. Technically, yield stress is defined as minimum stress required to initiate viscous flow.[51,52] At stresses below yield stress, the material does not flow but respond elastically. Therefore, yield stress gives a measure of the strength of the material and its ability to resist flow. In the literature, several testing techniques are employed to characterize the yield stress.[52,53] Some of these methods include steady shear protocols such start-up shear experiment and creep while application of oscillatory shear can also be employed to determine yield stress.[54] Firstly, yield stress can be estimated from the same experiment which was conducted to determine the linear viscoelastic region wherein strain amplitude was increased at a constant frequency. There have been multiple ways to estimate yield stress from this curve.[53] One way of defining yield stress and yield strain include identifying the point of maximum in the viscous modulus, as shown by line C in Figure 2. Another common technique of identifying yield point is the intersection point of the low and high stress asymptotes on an elastic modulus versus stress/strain amplitude curve as shown depicted by gray lines in Figure 2. [53] It is important to note that the estimated yield stress value is dependent on the applied frequency and an overshoot in viscous modulus is not ubiquitous to all complex materials. However, oscillatory amplitude sweep is a simple and quick technique to estimate yield stress.

Another conventional test to estimate yield stress is start-up shear experiment wherein a constant shear rate is applied and the growth of stress is recorded.[55,56] Here again, yield stress can either be taken as the point of deviation of stress from linear response, or point of maxima in stress or as the equilibrium stress at high strain/time values.[55] A graphical representation of the material behavior upon subject to step shear rate is shown in Table 1. It should be noted that overshoot in stress is not always exhibited by the sample and the nature of stress growth curve depends on the material and the applied shear rate.

Creep experiments also leads to the estimation of yield stress.[52] In a creep experiment, a constant stress is applied and the evolution of strain (or creep compliance) is observed as a function of time. In order to determine yield stress, creep experiments are performed at multiple values of stresses. At low values of stress, creep compliance will have a constant value while at higher stress values, it will grow with time. The stress value at which the transition in creep compliance curve occurs from constant to flowing behavior is identified as the yield stress.[55] A typical response of a soft material to creep experiment is shown in Table 1. While creep experiments give a good estimate of evaluating yield stress, it is a time taking experiment where multiple experiments are to be performed and a priori knowledge of approximate yield stress is important to design the experiment. Furthermore, this experiment is not suitable for samples which is undergoing structural changes gradually over time.

It is important to note that there exist several other methods of estimating yield stress. Yield stress is not a true material constant but depends on the employed measuring and analysis method. However, the value of yield stress estimated from all the techniques and nomenclatures is not significantly different.[52]

### Viscosity

In order to characterize the flow behavior of a sample, the most common rheological parameter is viscosity which can be measured by shear rate sweep technique. Most of the soft materials we commonly encounter are non-Newtonian in nature as a result, the viscosity changes as a function of applied shear rate. Furthermore, characterizing the viscosity of material as a function of shear rate helps in ensuring that products have the right texture, consistency, and flow properties for their intended use. In this experiment, a range of shear rate is applied in stepwise manner with defined measurement duration assigned to each point and response is measured in terms of viscosity and shear stress. It is important to choose the range of shear rate values correctly. At very low shear rates, it is difficult for the instrument to measure the weak torque signal. On the other hand, at high shear rates, turbulent flow effects within the gap could interfere with the flow, causing disturbances. One important but often overlooked parameter of rheological protocol is the acquisition time of each data point. This is especially critical in steady shear experiments, where the material needs enough time to reach a true steady-state flow, particularly at low shear rates. If the acquisition time is too short, the measurements can be inaccurate or misleading, which is a common issue when rheological techniques are used without careful consideration of these experimental details. In order to obtain accurate results, it is therefore important to either ensure to select steady state sensing option which is present in the rheometer software while designing the measurement protocol. Another approach is to set a fixed minimum acquisition time per data point in order to ensure steady state. A common guideline is to use a time duration that is a few multiple of time corresponding to the inverse of the lowest shear rate applied, which allows sufficient time for the material to respond fully at each measurement step. A typical range of shear rate is mentioned in Table 1, however, we can fine tune the range based upon the specific material under investigation and what shear rate does it gets exposed to during the processing condition. For instance, a new formulation of paint sample should be studied in the shear rate range of 0.001 $s^{-1}$ to 0.1 $s^{-1}$ while any material for dip coating application should be investigated in the range of 10 $s^{-1}$ to 100 $s^{-1}$.[57] Therefore, based upon the end use, the range of shear rate can be decided so that we have an actual understanding of the variation in viscosity of the sample when exposed to different processing conditions.

The typical response to a shear rate sweep in shown in Figure 9 which depicts the variation in viscosity as a function of shear rate. This information is particularly useful in understanding the flow behavior of the material under various shear conditions essential for processing and formulation design conditions. The variation in viscosity with shear rate helps in categorizing the flow behavior of the sample as Newtonian, shear thinning or thickening. For instance, in Figure 9, the viscosity curves of polymer solution and nanoclay dispersion corresponds to a shear thinning material wherein the material tends to flow easily when subjected to higher shear strain. Such shear thinning behavior is utilized in various applications including paints, food products, cosmetics, and 3D printing. On the other hand, we have shear thickening materials, as shown by silica dispersion, wherein the viscosity of the material increases with increase in shear strain. This shear thickening behavior is particularly beneficial for applications such as shock absorption in safety gear and body armour. For some materials, we observe constant value of viscosity in the low shear rate region which is referred as the zero-shear viscosity of the sample. Upon increasing the shear rate, the viscosity value decreases suggesting that the material flows easily with applied shear.

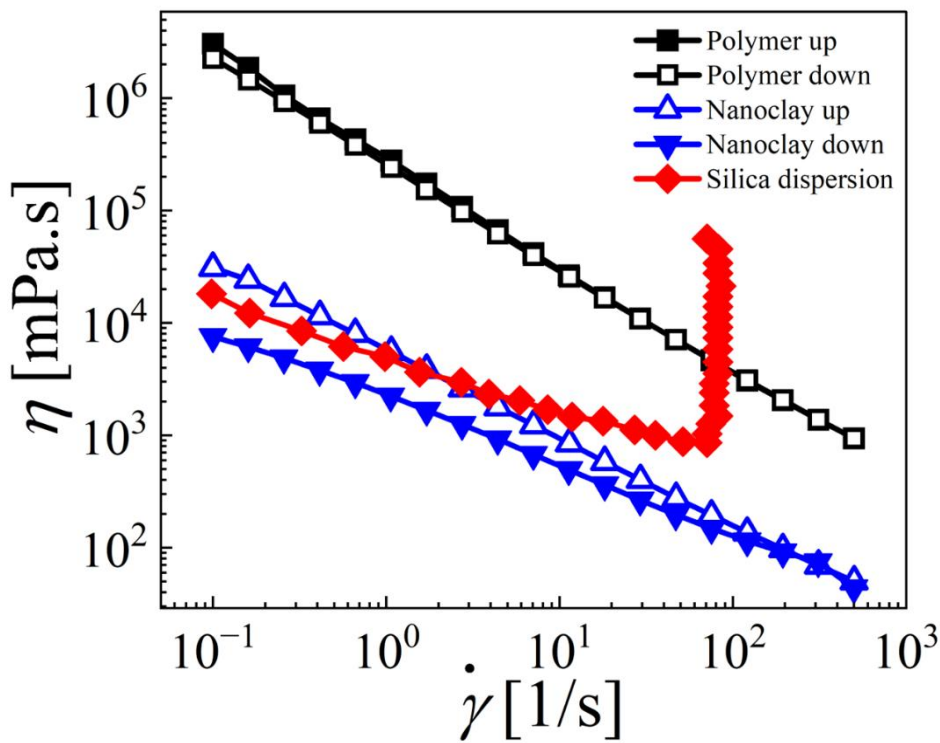

**Figure 9.** The variation in viscosity ($\eta$) is plotted as function of shear rate for a triblock copolymer solution (squares), nanoclay dispersion (triangles) and silica particles dispersion (diamonds, data taken from Ref.[6] ). The closed symbols represent that the shear rate was changed in an increasing (up) order while open symbols denote the shear rate was achieved in decreasing (down) order.

Furthermore, a shear rate sweep can be performed in increasing order of shear rate values (sweep up) and decreasing (sweep down) manner as well. Such experiment gives insights into the material response characteristics such as structural recovery or breakdown.[58] In Figure 9, we perform shear rate up and down experiment on two materials. It can be seen that the polymer gel does not exhibit any observable hysteresis with viscosity remaining almost identical when the shear rate is achieved either in increasing or decreasing order. However, the nanoparticle dispersion exhibits hysteresis loop in viscosity which suggests that the viscosity depends on the manner in which the shear rate is varied. The origin of hysteresis loop can be attributed to multiple factors such as thixotropy and shear banding.[59] The observed hysteresis in viscosity curves, the difference in the curves can indicate structural changes within the material, such as breaking or reformation of microstructures.[58,59] Therefore, shear rate sweep gives detailed information about viscosity and shear-dependent behavior, which helps in material characterization, process optimization, quality control, and formulation design in various industries.

All the rheological experiments provide insight about the material behaviour under different conditions. We learn about the response of the material as it evolves with time, frequency, temperature and shear rate. In Table 1, we present a comprehensive outline of all the fundamental rheological experiments with a typical operating range of input parameters, input protocol and output parameters. The details of each experiment in Table 1 provides a good foundation to design rheological experiments, obtain reliable data and interpret the obtained data effectively.

Stress Relaxation

In order to understand about the time dependent relaxation process in a material, a stress relaxation experiment can be performed. A constant magnitude of sudden strain is applied to the material and the subsequent decay in stress is recorded over time. At the moment of application of strain, most materials exhibit an immediate elastic stress. With increase in time, the stress gradually decreases as the material dissipates energy through molecular motion, rearrangement or flow. The long-time behavior is material dependent and the stress may either relax completely (long-term viscous behavior) or reach a plateau (long-term elastic behavior). The resulting stress relaxation modulus gives information about the relaxation mechanisms ranging from rapid molecular motions to slower network rearrangements. The obtained response can be fit to generalized Maxwell model to construct a relaxation time spectrum that gives information about the distribution of relaxation times in the material. Furthermore, stress relaxation experiment can also be helping in understanding the nonlinear behavior of the material.[60] Within the linear viscoelastic region, the magnitude of the applied constant strain does not influence the measured stress relaxation modulus. However, upon application of a constant strain magnitude which is beyond the linear viscoelastic region, the value of stress relaxation modulus is no longer independent of the magnitude of applied strain.[60] The stress relaxation curves decrease in magnitude upon increasing magnitude of deformation for a shear-thinning material. The amount of shifting in these curves can be utilized to model the deviation from linearity behavior of the material.[61] The stress relaxation experiment can be performed on a wide range of materials, including polymers, gels, biological tissues, and composites, informing both fundamental understanding and practical applications related to material design, processing, and performance.

**Table 1.** A comprehensive guide to design of rheological experiments. The table includes details of input parameters, output parameter, typical graphical representation of input function, output function and inference from the obtained rheological data.

| Experiment type | Inputs | Variation | Typical Value | Remarks | Graphical representation of Input | Graphical representation of Output | Inference from the experiment |
|---|---|---|---|---|---|---|---|
| **Linear Viscoelastic Region Analysis** | Strain [%]/stress [Pa] | Sweep | 0.1 – 1000 or 0.1 – 500 | Either strain or stress can be applied, Logarithmic increase in input | 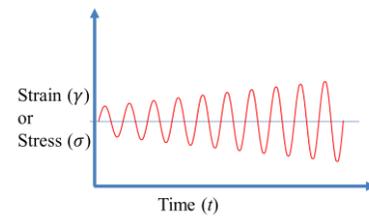 | 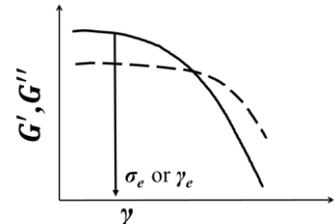 | Identification of linear domain of viscoelasticity wherein material properties are independent of applied deformation. |
| | Frequency [rad/s] | Constant | 1 | | | | |
| | Temperature [ºC] | Constant | 25 | Can be performed at temperature of interest | | | |
| | Time [s] | Decided by number of datapoints collected per decade | 5 datapoints collected per decade of deformation | | | | |
| | | | | | | | |

| **Frequency Sweep** | Strain [%]/stress [Pa] | Constant | $\gamma_e$ | The value of $\gamma_e$ obtained from amplitude sweep | 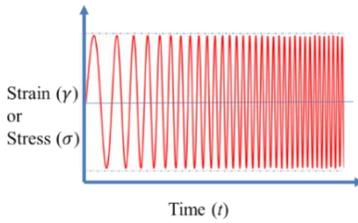 | 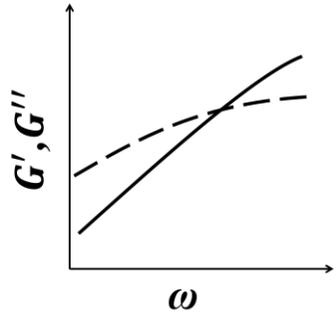 | • Dominance of elastic or viscous modulus indicative of material state<br>• $G'' \gg G'$: Liquid-like material behavior<br>• $G'' \ll G'$: Soft solid-like material behavior<br>• Crossover point indicative of relaxation time of the material |
|---|---|---|---|---|---|---|---|
| | Frequency [rad/s] | Sweep | 0.1 - 50 | Logarithmic increase, too high frequency can lead to instrument inertia interfering with the data | | | |
| | Temperature [ºC] | Constant | 25 | Can be performed at any temperature of interest | | | |

| | Time [s] | Decided by number of datapoints collected per decade | Typically 5 datapoints collected per decade of frequency | Number density of datapoints can be increased as per requirement | | | |
|---|---|---|---|---|---|---|---|
| **Aging** | Strain [%]/stress [Pa] | Constant | $\gamma_e$ | The value of $\gamma_e$ obtained from amplitude sweep | 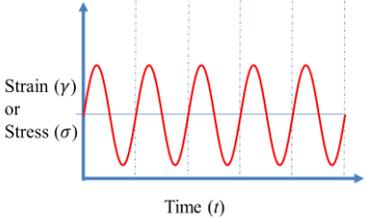 | 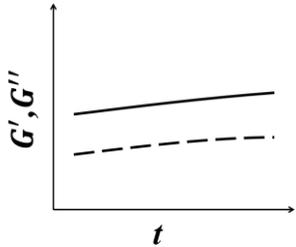 | • Indicative of stability of the dispersion<br>• Any abrupt changes in the rheological properties is indicative of instabilities |
| | Frequency [rad/s] | Constant | 1 | | | | |
| | Temperature [ºC] | Constant | 25 | Can be performed at any temperature of interest | | | |
| | Time [s] | Constant | 3600 | Can be conducted for longer duration as well | | | |

| **Temperature ramp** | Strain [%]/stress [Pa] | Constant | $\gamma_e$ | The value of $\gamma_e$ obtained from amplitude sweep | 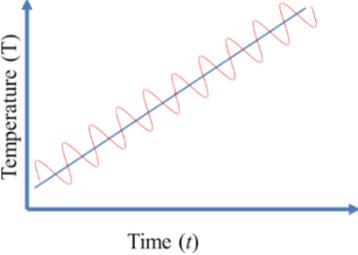 | 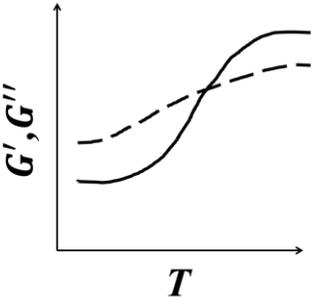 | • Indicative of effect of temperature on the material behavior, important test for temperature sensitive materials<br>• Depending upon the interparticle interaction and chemistry of the material, the rheological properties may increase, decrease or remain unchanged with temperature. |
|---|---|---|---|---|---|---|---|
| | Frequency [rad/s] | Constant | 1 | | | | |
| | Temperature [ºC] | Linear Ramp | 10 - 50 | Can be performed at any temperature range of interest either in heating or cooling manner, | | | |
| | | | | Ramp rate value can be provided | | | |
| | Time [s] | | | Will be decided by the ramp rate and range of temperature to be explored | | | |
| | | | | | | | |
| **Large amplitude oscillatory shear (LAOS)** | Strain [%] | Sweep | 0.1– 1000 | Logarithmic increase in input strain, Enable data recording for higher harmonics | 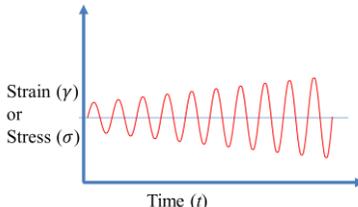 | 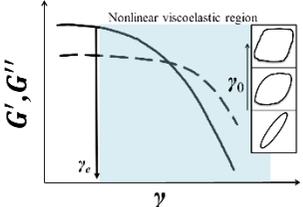 | Indicative of material behavior upon application of large amplitudes of deformations. |
| | Frequency [rad/s] | Constant | 1 | | | | |
| | Temperature [ºC] | Constant | 25 | | | | |
| | Time [s] | Decided by number of datapoints | 5 datapoints collected per | | | | |

| | | collected per decade | decade of deformation | | | | |
|---|---|---|---|---|---|---|---|
| | | | | | | | |
| **Yield stress** | | | | | | | Indicative of minimum stress required for the material to flow |
| Oscillatory amplitude sweep | Strain [%]/stress [Pa] | Sweep | | Performed in same way as Linear viscoelastic region analysis | 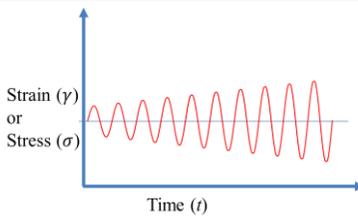 | 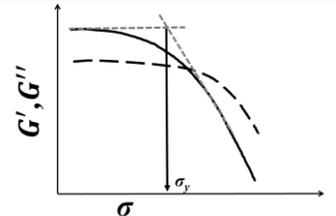 | Yield point can be defined as:<br>• intersection point of the low and high amplitude asymptotes, or<br>• maximum in $G''$. |
| Start-up shear | Shear rate [s$^{-1}$] | Constant | 1 | Can be performed at any other constant shear rate values | 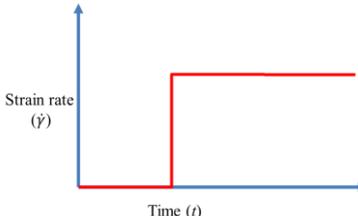 | 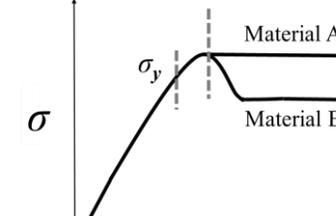 | Yield point can be defined as:<br>• point of deviation of stress from linear response, or<br>• point of maxima in stress or<br>• as the equilibrium stress at high strain/time values |

| Creep | Stress | Constant | Any value | Need to conduct multiple such experiments at different stress values each time, A priori idea of range of stress where yield stress can lie is helpful | 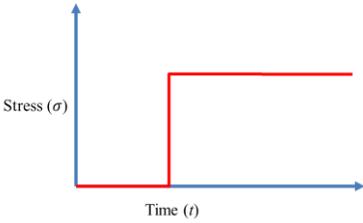 | 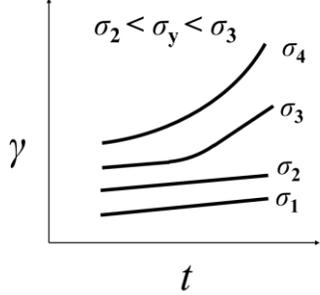 | Yield stress is the creep stress at which viscous (flowing) behavior is observed |
|---|---|---|---|---|---|---|---|
| | | | | | | | |
| **Viscosity** | Shear rate [s$^{-1}$] | Sweep | 0.1-500 | Logarithmic increase, Caution about too high shear rate value can lead to spillage of sample, Ensure to check the box to have | 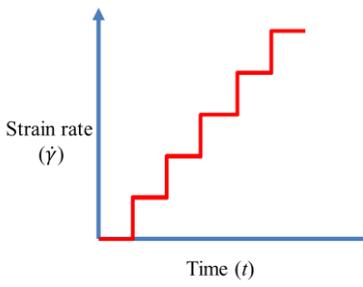 | 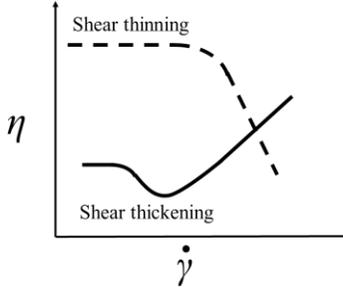 | • Gives measurement of viscosity of the material at varying shear rates<br>• Indicative of shear thinning or thickening behavior |
| | | | | steady state | | | |
| | Temperature [ºC] | Constant | 25 | Can be performed at any temperature of interest | | | |
| | Time [s] | | | Will be decided by the shear rate range explored | | | |
| | | | | | | | |
| **Stress Relaxation** | Strain [%] | Constant | Preferably at magnitude within LVE ($\gamma < \gamma_e$) | Can be performed at any temperature of interest | 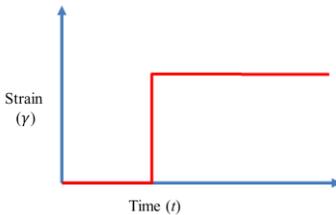 | 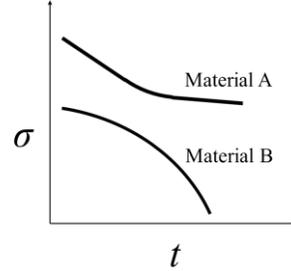 | • Gives information about the relaxation time spectrum<br>• Material A: long-term elastic behavior (reaches a plateau)<br>• Material B: long-term viscous behavior (relaxes completely) |

## Best practices

We have discussed a variety of rheological experiments and their importance in understanding the behavior of the material. However, it is equally important to ensure that the obtained results are free of any measurement artifacts. These artifacts may arise due to multiple factors such as preconditioning, sample volume and wall slippage. Such artifacts can interfere with the true response of the material behavior and lead to inaccurate conclusions. We explore methods for detecting and minimizing such experimental errors to ensure the accuracy and reliability of rheological data. By addressing potential sources of error, such as sample inconsistencies, wall slip, sample volume, we provide strategies to reduce their impact. It is important to carefully identify the experimental challenges and take appropriate measurement to achieve precise results and improve the overall quality of rheological experiments.

### Consistency in sample preparation

In order to get accurate and reliable rheological data, it is of utmost importance that the sample preparation step remains identical each time. It is crucial to ensure that all sample preparation parameters such as purity of material, stirring speed, drying, pH, temperature have been kept constant across all batches of sample preparation. Any change in any one of the preparation protocol step can lead to changes in the rheological behavior of the materials. Recently, Petekidis and coworkers highlighted the role of stirring temperature while sample preparation on the rheological properties of the formed organoclay gels.[62] Importantly, the sample should be free of any trapped air bubbles as it can lead to erroneous results. We can use sonication or any other degassing technique to ensure the sample to be devoid of any air bubble.

### Choice of geometry

When selecting a rheometer geometry, it is essential to consider the nature of the sample, the type of measurement desired, and the rheological properties of interest. The common rheometer geometries include parallel plate, cone and plate and concentric cylinder geometries, each suited to different sample types and shear conditions. While all the varieties of geometries are compatible for all kinds of rheological measurements, one can consider few points while selecting the geometry. The concentric cylinder geometries are commonly employed for samples that maybe unstable and to minimize the evaporation losses during the measurement. However, this geometry is best suited for weakly structured materials, as samples containing strong aggregates, such as thermosetting samples and cementitious materials, can pose challenges during sample removal due to aggregation. The cone and plate geometry is usually used for small sample volumes, normal stress measurements and provides a uniform shear rate across the geometry. Furthermore, the parallel plate geometry is the most versatile type of geometry which can be used for particulate samples and thermosetting materials and allows for the study of materials under different gap settings. It is important to note that all these types of geometries come in varying diameters and cone angles to accommodate a wide range of materials. For less viscous samples, selecting a larger geometry dimension is advisable to achieve optimal measurement signals, as this increases the sample volume and enhances the sensitivity and accuracy of the rheological data.

### Preconditioning of the sample

Pretreatment of the sample is performed to erase any history of the sample so that we get a standard initial state before the beginning of any experiment. This reduces variability and helps ensure accurate and reproducible results. Pretreatment process can be done in either in oscillatory or steady mode. In oscillatory mode, usually a large strain amplitude (much greater than $\gamma_e$) is applied at a constant frequency. In steady mode, we can apply a fairly large amount of constant shear rate for a fixed duration of time such that any pre-existing microstructure in the sample is broken. It is also important to equilibrate the sample after loading which can be done by allowing the sample to rest for a predetermined duration at a constant temperature. This is especially important for samples that are temperature or shear sensitive. Therefore, preconditioning step is important to have the same initial state before conducting any measurement.[63,64]

### Optimum Sample loading

It is crucial to ensure optimal sample loading, particularly when using cone-plate or plate-plate geometries. Improper sample loading can lead to measurement inaccuracies, where overfilling the sample results in elevated values of the measured parameter, while under-filling may cause a reduction in the observed measurement values. Ewoldt and coworkers showed that slight overfill or underfill can disrupt the rotational symmetry and surface tension affects can affect the rheological measurements leading to erroneous results.[23] Therefore, care must be taken to trim the sample when prompted by the instrument. For accurate measurements, the sample should be of a volume that ensures a uniform distribution between the measuring surfaces, and minimizing errors due to sample inconsistency. For samples containing solvents which can evaporate, it is important to apply a thin layer of low viscosity oil to prevent the sample from drying out. Additionally, we can also use an additional hood or solvent trap.

### Mutation number

It is very common for a complex material to continuously evolve as a function of time. Under such scenario of continuously evolving system, it becomes important to understand how to obtain material properties at a given frequency sweep. In principle, we would like to have knowledge of moduli over a wide range of frequency while conducting a frequency sweep experiment. However, for an evolving system, we need to carefully select the frequency domain such that the sample is not mutating (changing) significantly within the interval time of one frequency sweep. To solve this, it is advisable to compute the mutation number ($N_{mu}$) defined as:[36]

$$N_{mu} = \Delta t \frac{1}{g} \frac{\partial g}{\partial t}. \tag{1}$$

The mutation number gives an estimate of the relative change in measured parameter $g$ over the sampling period of $\Delta t$. It can be calculated from the value of elastic and viscous modulus as:[36]

$$N'_{mu} = \frac{2\pi}{\omega G'} \frac{\partial G'}{\partial t} \quad \text{and} \quad N''_{mu} = \frac{2\pi}{\omega G''} \frac{\partial G''}{\partial t}. \tag{2}$$

The change in mechanical property of a material during an experimental time window of $2\pi/\omega$ is considered negligible if $N_{mu}$ is below an acceptable tolerance level of 10% [36].

Below this threshold value, the material can be regarded as quasi-stable during the experimental duration. Under this scenario, every measured value of $G'$ and $G''$ corresponds to a unique quasi-stable state of the material. Therefore, it is advisable to compute mutation number for an evolving sample to help ensure measurement accuracy and optimize process parameters.[36] Only those datasets which obeys the rule of $N_{mu} < 0.1$ should be considered as true data depicting the material behavior.

## Wall Slippage

It is usually assumed that the sample loaded in the shear cell adheres to the surface of the geometry boundaries and is known as the no-slip condition. However, wall slip is a common phenomenon observed in flow viscosity measurements or in samples containing oil, fat and biofluids.[65,66] The high velocity gradients in a thin layer near the geometry wall leads to wall slip. Experimentally, wall slip manifests in lower measured value of viscosity than the true viscosity of the sample, or early shear thinning or lower value of yield stress. One way to diagnose the wall slip effect in the sample is to conduct measurement at different gap thicknesses. If the material property is independent of gap thickness, then there is no wall slip in the sample.[23] In case of wall slip, one can minimize the wall slip effect by using serrated, roughened, crosshatched or vane type geometries. In case of unavailability of roughened geometries, a practical and effective workaround is to modify existing smooth surface geometries by attaching sandpaper to them. This can be done by securely gluing a piece of sandpaper onto the surface of the smooth geometry using a suitable adhesive.[67] The purpose of this modification is to introduce surface roughness, which helps in minimizing or eliminating wall slip during rheological measurements. After attaching the sandpaper, it is essential to perform a zero gap calibration. This step ensures that the added thickness of the sandpaper is accurately accounted for in the geometry's gap setting. The glued sandpaper on smooth geometry can then function as a roughened geometry which can be used to prevent wall slip. Therefore, preventing wall slippage in rheological measurements is essential for obtaining accurate data, especially when testing low-viscosity or shear-thinning materials.[68,69]

We outline some of the best practices to follow while performing the rheological experiments. However, there are various other experimental challenges that can arise such as edge fracture, secondary flows and surface tension effects. A thorough and insightful discussion of the experimental challenges encountered in rheological measurements is presented in the article by Ewoldt *et. al*.[23] Upon completion of the rheological experiment, the data from most tests including oscillatory shear, creep, and steady shear are plotted on double logarithmic axes. This type of plotting captures variations in rheological data without distortion, ensuring that both short-term and long-term material responses are appropriately visualized. Furthermore, using semi-log or linear axes can increase the probability of misinterpretations, such as underestimating terminal relaxation times or exaggerating relaxation rates.[21]

In the era of advanced materials, it is becoming increasingly important to standardize and list some of the fundamental rheological experiments which can be beneficial in material characterization. In this paper, we have presented a comprehensive framework of performing rheological experiments to gain understanding about the flow behavior of material and utilize the obtained information in optimising the product development. Through the series of proposed experiments, it is possible to have an understanding about the viscosity at varying deformations, stability with respect to time and temperature, modulus and yield stress of the material. These rheological parameters are critical in predicting the shelf life and performance of the material. Moreover, we also discuss the best practices to follow in order to avoid bad data and artifacts and propose techniques which can be utilized to identify such artifacts and avoid them.

## Advanced Rheology

This section highlights selected applications of advanced rheological techniques. This discussion, while not exhaustive, aims to highlight the various approaches through which advanced rheological measurements can be employed to gain deeper insights into material behavior. Rheological analysis can be enhanced by integrating it with multiple analytical techniques to provide a comprehensive understanding of material behavior. Microscopy techniques such as optical and confocal microscopy can be employed in-situ while performing rheological measurements which helps in flow visualization in real time.[70,71] We can also couple the rheological measurements with light scattering technique to get an insight about the particle size distribution.[72] Furthermore, rheometer can also be setup in the beamtime to perform Rheo-SAXS and Rheo-SANS measurements.[70,73] Such an arrangement probes the microstructural changes during shear which provides insights about the internal arrangement of microstructure. Furthermore, interfacial rheology can be performed to investigate the surface active materials and get insight about flow and deformation behavior at the interface between two phases. In addition to the techniques mentioned above, rheological measurements can be tailored in various innovative ways to suit specific applications such as thin-gap rheology, microrheology, or rheo-NMR, offering greater flexibility and insight across diverse research needs.[70,74,75] There also exists another class of colloidal suspension which respond to changes in electric and magnetic fields and find applications as shock absorbers, brakes and dampers. These materials can be characterized by coupling the rheometer with the magnetic and electric field generator and understanding the changes in the viscoelastic properties upon application of electric or magnetic field. [76,77] This coupling enables a deeper understanding of field-induced structural changes within the material, thereby informing the design of materials optimized for adaptive and responsive applications.

In many industrial processes such as extrusion, coating, and mixing, the material is subjected to multiple stress and strain fields simultaneously. In order to understand the behavior of the material under real-world processing conditions, there have been advanced rheological protocols. Few of such techniques includes application of oscillatory shear perpendicular to the primary steady shear flow. This method is called orthogonal superposition rheometry which enables measurement of the materials viscoelastic behavior under nonlinear flow field.[78] Furthermore, the oscillatory perturbation can also be applied in the same direction as the primary flow which is known as the parallel superposition rheometry. This method provides information about the flow-direction anisotropy in structured materials.[79] These advanced superposition-based rheological techniques aids in understanding the internal dynamics of the material under shear and offer a powerful way to bridge the gap between the fundamental tests and real-world process conditions.[80] Furthermore, many highly viscous materials such as polymeric melts and doughs can be characterized using squeeze-flow technique wherein the material is compressed between the parallel plates. This technique is particularly helpful in measurement of viscosity of highly viscous materials which is otherwise difficult to measure in conventional experiments due to operational issues such as melt and edge fracture.[81,82] Furthermore, there is also provision in some specialized rheometers to conduct experiments at very high frequency which enables probing of rapid dynamics, giving valuable information about the microstructure at the local scale and provides detailed insight into the interactions between hydrodynamic forces and interparticle effects.[83] High frequency rheometry can be utilized for studying the particle dynamics in shear thickening solutions at high shear rates, analyzing entanglement dynamics polymer melts and solutions at short timescales and investigating the fast dynamics in food and biological materials.[84,85] With this introductory knowledge of the advanced concepts in rheology, one can make informed decisions about which experimental techniques are most-suited for their material of interest and its intended applications and the details of these advanced rheological measurement can be found in the respective cited references.

## Knowledge Roadmap

In order to make rheology more accessible and inclusive, we offer a clear, concise pathway designed to support users at all experience levels. An understanding of these different levels will make it easier to guide the learning pathway and helps researchers build on what they know and elevate their rheological skills.

- *Beginners*: For those new to the field of rheology, one should focus on learning about the fundamentals of rheology, knowing the key rheological variables and understanding the basic working principle of rheometer. It is important to understand the machine variables such as torque and displacement to ensure proper instrument calibration and selecting appropriate geometries. The most crucial step is designing a basic rheological test. As shown in Figure 1, one of the first objective when performing rheological characterization should be identification of linear viscoelastic region. Once identified, one can perform frequency sweep, aging, viscosity measurements to characterize the key rheological properties of the material. The stability of the sample across varying temperature can be analyzed by performing a temperature ramp. Table 1 helps in designing these rheological experiments wherein a reference of the input and output variables are mentioned. Furthermore, a representative response of a viscoelastic material is presented in Table 1, followed by guidance on how to interpret the corresponding output data. Another important learning is to be aware of the common sources of errors and the best practices to follow to get accurate and reliable results.
- *Intermediate*: After gaining fundamental understanding about the viscoelastic behavior of the material, one can progress towards performing measurements to reveal deeper insights into the material behavior. These measurements include yield stress, LAOS and stress relaxation experiments. It is important to understand the various data analysis methods for LAOS and interpreting the nonlinear rheological parameters. Furthermore, analysis of the stress relaxation measurement result allows for understanding the various relaxation modes present in the viscoelastic material. These analyses provide valuable information on nonlinear viscoelastic behavior and can enable users to characterize complex material responses beyond the linear regime.

- *Advanced*: After characterizing the material's flow behavior through rheological analysis, it is essential to investigate the underlying microstructural changes that drive the observed bulk properties. A comprehensive understanding of microstructural changes can be achieved by integrating rheological measurements with complementary analytical techniques such as optical microscopy and scattering. Furthermore, there are more complex materials which are either evolving with time or stimuli-responsive. Such stimuli-responsive behavior can be studied by integrating rheometer with an electric/magnetic field generator. Also given the widespread presence of complex materials across various industries, it is important to understand their flow behavior under real-world processing conditions. These complex processing conditions can be replicated by applying oscillatory flow field with the steady shear field using superposition rheology. Furthermore, based on the specific application, one can also perform thin-gap rheology, squeeze flow rheometry, or high-frequency rheology to gain an enhanced understanding of the material's behavior under conditions that closely mimic real-world processing or operational environments. These specialized techniques provide insights into phenomena such as thin film flow, compression and deformation between surfaces, and rapid dynamic responses, respectively, thereby offering a more comprehensive characterization of complex materials. By combining these approaches, researchers can correlate macroscopic flow behavior with underlying microstructural dynamics, enabling a more thorough characterization of complex materials.

## Summary

We encourage all material scientists to employ rheology as a vital characterization technique to gain insight about the mechanical property of the material. The focus of this work has been on understanding the most common rheological experiments and estimating the key rheological parameters which is crucial for effectively characterizing the material's flow and deformation behavior. This article provides a roadmap of rheological experiments aiming to guide all researchers in designing the experiments tailored to their specific needs and intended applications. The experiments discussed here provide a foundational understanding of the soft material's mechanical response, which is vital for designing products that possess the desired viscosity, modulus, and stability. By conducting these initial rheological tests, researchers can gain critical data that will inform the development of materials with specific performance characteristics. Furthermore, based on the insights gained from the initial experiments outlined in this work, additional, more advanced rheological tests can be planned. These follow-up experiments will allow researchers to refine their understanding of material behavior and design more precise and optimized products for various industrial applications.

This work also highlights the best practices to be employed to get accurate rheological data. The article also highlights the best practices that should be employed to obtain accurate and reliable rheological data. Ensuring precise measurements is critical for drawing meaningful conclusions about material behavior. These best practices include proper calibration of rheological instruments, careful sample preparation, and controlling environmental factors such as temperature and humidity during experiments. Ultimately, rheology offers a powerful tool for material scientists, enabling them to fine tune the mechanical properties of materials to meet the demands of specific processes and end uses.

## Acknowledgements
The author acknowledges the financial support from Anusandhan National Research Foundation (ANRF), Government of India (ANRF/ECRG/2024/001771/ENS) and startup research grant from IIT Madras.